\documentclass[prd,aps,preprint,tightenlines,superscriptaddress,nofootinbib,showpacs,showkeys,floatfix]{revtex4}
\usepackage{mathrsfs}
\usepackage{amsfonts}
\usepackage{array}
\usepackage{epsfig}
\usepackage{rotating}
\usepackage{multirow}

\usepackage{amsmath}    
\usepackage{verbatim}   
\usepackage{color}      
\usepackage{colordvi}

\usepackage{subfigure}  
\usepackage{hyperref}   

\usepackage{pstricks,rotating, graphicx}
\usepackage{cancel}

\usepackage{tablefootnote}  

\graphicspath{{./Pics/}}
\DeclareGraphicsExtensions{.pdf}

\usepackage{pslatex}
\usepackage{txfonts}
\usepackage[latin1]{inputenc}
\usepackage[T1]{fontenc}

\begin{document}

\begin{titlepage}
\thispagestyle{empty}
\noindent
DESY 17-114\\
DO-TH 17/17\\
\hfill
August 2017 \\
\vspace{1.0cm}

\begin{center}
  {\bf \Large
    Strange sea determination from collider data
  }
  \vspace{1.25cm}

 {\large
   S.~Alekhin$^{\, a,b}$,
   J.~Bl\"umlein$^{\, c}$,
   and
   S.~Moch$^{\, a}$
   \\
 }
 \vspace{1.25cm}
 {\it
   $^a$ II. Institut f\"ur Theoretische Physik, Universit\"at Hamburg \\
   Luruper Chaussee 149, D--22761 Hamburg, Germany \\
   \vspace{0.2cm}
   $^b$Institute for High Energy Physics \\
   142281 Protvino, Moscow region, Russia\\
   \vspace{0.2cm}
   $^c$Deutsches Elektronensynchrotron DESY \\
   Platanenallee 6, D--15738 Zeuthen, Germany \\
 }
  \vspace{1.4cm}
  \large {\bf Abstract}
  \vspace{-0.2cm}
\end{center}
    We consider determinations of the strange sea in the nucleon based on 
    QCD analyses of data collected at the LHC with focus on the  
    recent high-statistics ATLAS measurement of the $W^\pm$- and $Z$-boson production.
%
%
   We study the effect of different functional forms for parameterization of
   the parton distribution functions and the combination of various data sets 
   in the analysis.
   We compare to earlier strange sea determinations and discuss ways to
   improve them in the future.
\end{titlepage}

\subsection*{1.  Introduction}
The precise knowledge of the light-quark content of proton is very important for
phenomenological studies at the Large Hadron Collider (LHC). 
QCD analyses of data from colliders and fixed-target experiments make 
this information available through the parton distribution functions (PDFs), 
which nowadays are accurate to next-to-next-to-leading order (NNLO) in
perturbation theory~\cite{Accardi:2016ndt}.
It has been shown~\cite{Alekhin:2017kpj}, that the recent LHC data on $W^\pm$- and $Z$-boson production 
provide valuable constraints on the light-quark distributions for up, down and strange 
and help to improve the flavor separation.
Currently, however, the extraction of the strange sea carries the largest uncertainty, which, for instance, 
plays a crucial role in the precision of the recent $M_W$-mass measurement by the ATLAS experiment~\cite{Aaboud:2017svj}.
It is therefore of particular importance to pin down the strange sea determination to better accuracy.

Such an improvement can be achieved with the $W^\pm \to l^\pm \nu$ and $Z \to l^+l^-$ cross section measurements 
of the ATLAS experiment~\cite{Aaboud:2016btc}. 
However, the ATLAS analysis, which has been published as the so-called epWZ16 set of PDFs~\cite{Aaboud:2016btc}, 
has obtained a strange-quark sea of a size comparable to the non-strange-quark ones  
in the kinematic range of Bjorken $x\sim 0.01$.
In this way, ATLAS has confirmed with better accuracy 
its earlier results~\cite{Aad:2012sb} based on a smaller data sample~\cite{Aad:2011dm}.
An enhancement of the strange-sea was observed by ATLAS 
also in an analysis of its data on the associated production of 
$W^\pm$-bosons and a charm-quark~\cite{Aad:2014xca}, 
which were well described by its epWZ12 PDF set published in Ref.~\cite{Aad:2012sb}.
On the other hand, an analysis of the $W^\pm$+charm data collected by the CMS
experiment~\cite{Chatrchyan:2013uja} does not show any such strange sea
enhancement. 

In a wider context, this situation is problematic, because the ATLAS results
also disagree with the strange sea PDFs extracted from other processes.
First of all, there is data on charm-quark production in the neutrino-induced
deep-inelastic scattering (DIS) off nucleons.  
This process, initially measured with a good accuracy by the CCFR and NuTeV
experiments at Tevatron~\cite{Goncharov:2001qe}, 
was later studied with an even better accuracy by the NOMAD experiment at 
CERN's SPS collider~\cite{Samoylov:2013xoa}.
All three experiments prefer a stronger suppression of the strange sea 
as compared to the ATLAS one~\cite{Alekhin:2014sya}.
Moreover, the ATLAS findings of an almost perfect flavor ${\rm SU(3)}$ symmetry 
among the three light sea quark distributions have not been confirmed 
in global fits of PDFs, as reviewed for instance in~\cite{Accardi:2016ndt}.

The present paper aims at clarifying these discrepancies and at consolidating
the different strange sea determinations. For this purpose we use  
the global ABMP16 PDF fit~\cite{Alekhin:2017kpj} as a framework. 
We consider variants of the ABMP16 fit with different shapes for the
functional form of the PDF parameterization at the initial scale of the fit 
as used by ATLAS and in the ABMP16 analyses.
We also consider combinations of different sets of data from colliders and fixed-target experiments.   
In this way we can separate the impact of different effects on the strange sea
determination and localize the origin of discrepancies.

\subsection*{2. Shape of PDF parameterizations}
The ABMP16 analysis~\cite{Alekhin:2017kpj} is performed at NNLO accuracy in QCD and 
the PDF extraction is based on inclusive DIS and Drell-Yan (DY) 
data supplemented by data on the DIS- and hadro-production of heavy quarks.
In the ABMP16 fit, the PDFs are parameterized at a starting scale
$\mu_0^2=9~{\rm GeV}^2$ for the QCD evolution in a scheme with $N_F=3$ light
flavors as follows 
\begin{eqnarray}
\label{eq:abmp16-pdfs}
  x q_{v}(x,\mu_0^2)
  &=&
  \displaystyle
  \frac{ 2 \delta_{qu}+\delta_{qd}} {N^{v}_q}\, (1-x)^{b_{qv}}\, x^{a_{qv}\, P_{qv}(x)}
  \, ,
  \nonumber\\
  xq_{s}(x,\mu_0^2) 
  &=&
  A_{qs}\, (1-x)^{b_{qs}}\, x^{a_{qs}\, P_{qs}(x)}
  \, ,
  \nonumber\\
  xg(x,\mu_0^2)
  &=&
  A_{g}\, (1-x)^{b_{g}}\,  x^{a_{g}\, P_{g}(x) }
  \, ,
\end{eqnarray}
with the valence quark distributions ($q_{v}$ for $q = u,d$), 
the sea quark distributions ($q_{s}$ for $q = u,d,s$), 
assuming $q_s(x,\mu_0^2) = \bar{q}_s(x,\mu_0^2)$, 
and the gluon. 
The functional form of the PDFs is controlled by 
the exponents $a_{p}$ and $b_{p}$ and the functions $P_{p}(x)$ of the form
\begin{equation}
\label{eq:abmp16-poly}
P_{p}(x) \,=\, (1+\gamma_{-1,p}\ln x)
\left( 1 + \gamma_{1,p}x + \gamma_{2,p}x^2 + \gamma_{3,p}x^3 \right)\, ,
\end{equation}
where $p=qv,qs,g$.
The normalizations $N_q^v$ and $A_g$ in Eq.~(\ref{eq:abmp16-pdfs}) 
have been determined from the fermion number and momentum conservation sum
rules and $\delta_{qq^\prime}$ denotes the Kronecker symbol.
All other 25 parameters $A_{qs}, a_{p}, b_{p}$ and $\gamma_{p}$ are fitted to the data.
It has been checked in the ABMP16 fit, that Eq.~(\ref{eq:abmp16-poly}) 
allows for sufficient flexibility of the PDFs in the entire range of Bjorken-$x$
covered by the data which are included into the fit.

In contrast, the analysis of the ATLAS $W^\pm \to l^\pm \nu$ and $Z \to l^+l^-$ 
cross section measurements for the extraction of the epWZ16 PDFs~\cite{Aaboud:2016btc} 
at NNLO in QCD has used a much restricted set of data. 
ATLAS only includes its own data on $W^\pm$- and $Z$-production in combination 
with DIS data from the HERA collider. 
The epWZ16 PDFs are derived from the following parameterizations at the
starting scale $\mu_0^2=1.9~{\rm GeV}^2$, 
\begin{eqnarray}
  x u_\mathrm{v}(x,\mu_0^2) &=&  A_{u_\mathrm{v}} x^{B_{u_\mathrm{v}}} (1-x)^{C_{u_\mathrm{v}}} ( 1 + E_{u_\mathrm{v}} x^2)\, ,\nonumber \\
  x d_\mathrm{v}(x,\mu_0^2) &=&  A_{d_\mathrm{v}} x^{B_{d_\mathrm{v}}} (1-x)^{C_{d_\mathrm{v}}}              \, ,\nonumber \\
  x \bar{u} (x,\mu_0^2) &=& A_{\bar{u}} x^{B_{\bar{u}}} (1-x)^{C_{\bar{u}}}\, , \nonumber \\  
  x \bar{d} (x,\mu_0^2) &=& A_{\bar{d}} x^{B_{\bar{d}}} (1-x)^{C_{\bar{d}}}\, , \nonumber \\
  x g(x,\mu_0^2)   &=& A_g x^{B_g} (1-x)^{C_g} - A'_gx^{B'_g}(1-x)^{C'_g}\, , \nonumber \\
  x \bar{s}(x,\mu_0^2) &=& A_{\bar{s}} x^{B_{\bar{s}}} (1-x)^{C_{\bar{s}}} \, ,  
\label{eq:epWZ16-pdfs}
\end{eqnarray}  
for $N_F=3$ light flavors, and assuming $s = \bar{s}$ for the strange sea.
Again, the sum rules for fermion number and momentum conservation determine
the normalizations $A_g$, $A_{u_\mathrm{v}}$ and $A_{d_\mathrm{v}}$.
In addition, the parameter $C'_g=25$ is fixed by hand to a large value and 
the assumption of iso-spin symmetry at small $x$, i.e., $\bar u=\bar d$ as $x\to 0$
is invoked to set $A_{\bar{u}} = A_{\bar{d}}$ and $B_{\bar{u}} = B_{\bar{d}}$.
Finally, the strange sea at small $x$ is assumed to be related to the light
quark sea, $\bar{u}$ and $\bar{d}$, so that $B_{\bar{s}}=B_{\bar{d}}=B_{\bar{u}}$ is put by hand. For these assumptions there is neither theoretical evidence nor
are they indicated by fits using more general parameterizations,
as will be shown below.
This leaves a total of $15$ variables in Eq.~(\ref{eq:epWZ16-pdfs}) to be determined from data. 
The PDF shape Eq.~(\ref{eq:epWZ16-pdfs}) is motivated by the predecessors of the epWZ16 analysis, 
the PDF fits of HERAPDF family, which are based exclusively on HERA data 
and therefore have to impose several constraints on the PDF shapes 
in kinematic regions of Bjorken-$x$, which are not sufficiently covered by the HERA data.

\bigskip

In order to check the consistency of the assumptions underlying the epWZ16 PDFs 
with the data at large Bjorken-$x$, which commonly constrain the PDFs in global
analyses we perform a test variant of the ABMP16 fit using the PDF shapes 
of Eq.~(\ref{eq:epWZ16-pdfs}). 
In addition, the collider data from LHC and Tevatron on rapidity
distributions for $W^{\pm}$- and $Z$-boson production in the electron- and
muon-decay channels as well as for lepton-charge asymmetries (see Tab.~2 in \cite{Alekhin:2017kpj}) are replaced by
data for DIS off deuterons (see Tab.~3.2 in \cite{Alekhin:2012ig}). 
The latter had been omitted in the ABMP16 fit since they require taking into 
corrections for the nuclear effects in the deuteron target, which bring in 
an additional source of uncertainty. 
In the meantime, though, the PDFs extracted with deuteron DIS data included and
using the shape of deuteron corrections suggested by the off-shellness model 
of Kulagin-Petti~\cite{Kulagin:2004ie} 
have been shown to be in agreement with the ones preferred by the 
$W^{\pm}$- and $Z$-boson collider data~\cite{Alekhin:2017fpf}.
Therefore, the deuteron DIS data allow to obtain a reliable constraint 
on the light-quark PDFs ($u$, $d$) in the range $x\gtrsim 0.01$. 
This approach avoids the tedious computation of predictions for the 
lepton rapidity distributions in $W^{\pm}$- and $Z$-boson production 
with account of kinematic cuts by means of fully differential codes, like {\tt FEWZ} 
(version 3.1)~\cite{Li:2012wna,Gavin:2012sy}.
Thus, the use of deuteron DIS data leads to a fast and efficient fit,
since the relevant DIS cross sections are evaluated at NNLO in QCD with 
the code {\tt OPENQCDRAD} (version 2.1)~\cite{openqcdrad:2016}.

For the purpose of comparison we also consider the variant of fit 
with the same data selection and the 
ABMP16 shape for the PDFs in Eq.~(\ref{eq:abmp16-pdfs}).
The notations used throughout the paper to present results of these two
variants are the following:
\begin{itemize}
\item [ ] \textit{\textbf{ABMP16 shape}}
  -- 
 a fit referring to the $W^{\pm}$- and $Z$-boson collider data replaced by deuteron DIS data 
  and the PDF shape of Eq.~(\ref{eq:abmp16-pdfs}),

\item [ ] \textit{\textbf{epWZ16 shape}}
  -- the same but
  with the PDF shape of Eq.~(\ref{eq:epWZ16-pdfs}).
\end{itemize}
\begin{figure}[t!]
  \centerline{
    \includegraphics[width=7.85cm]{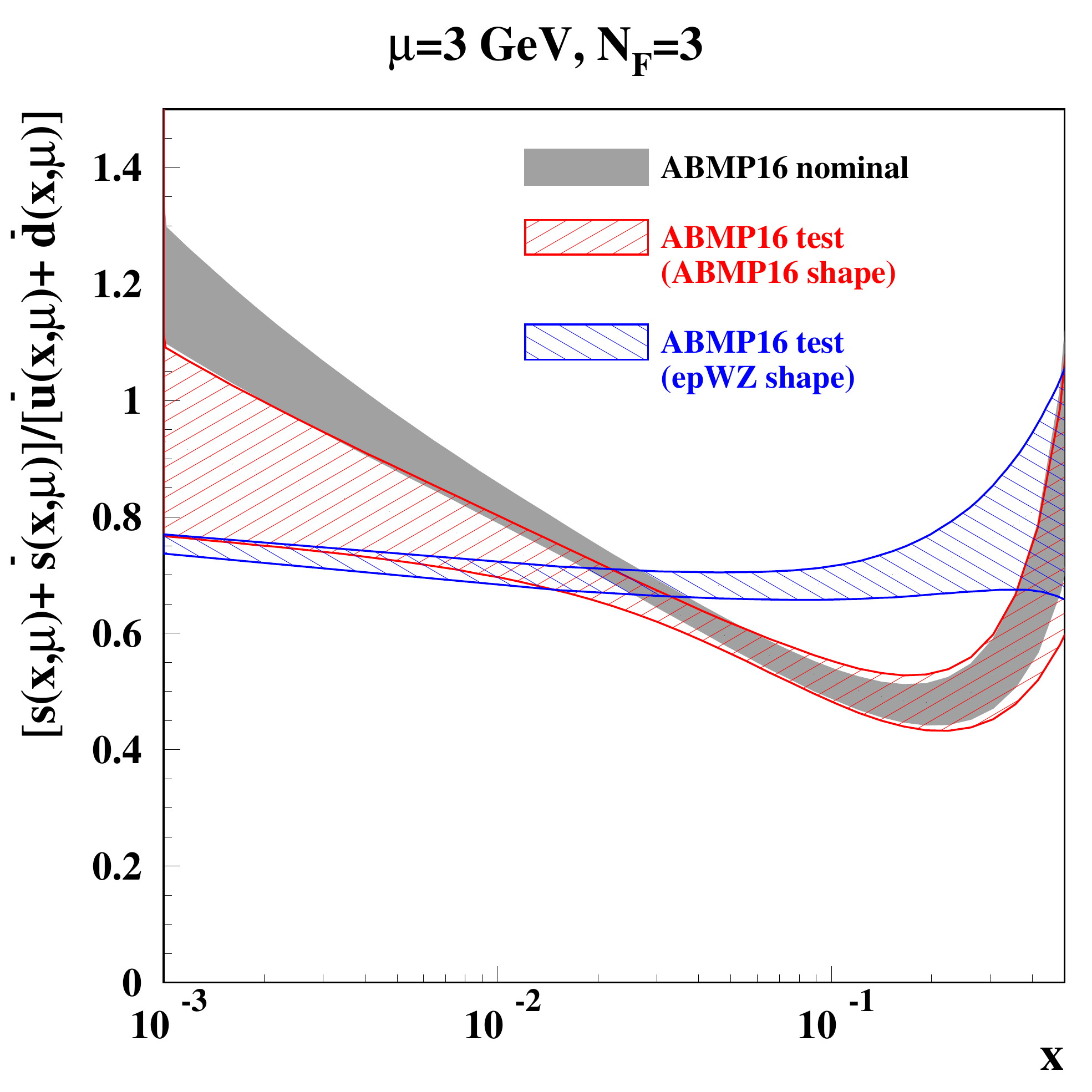}
    \includegraphics[width=7.85cm]{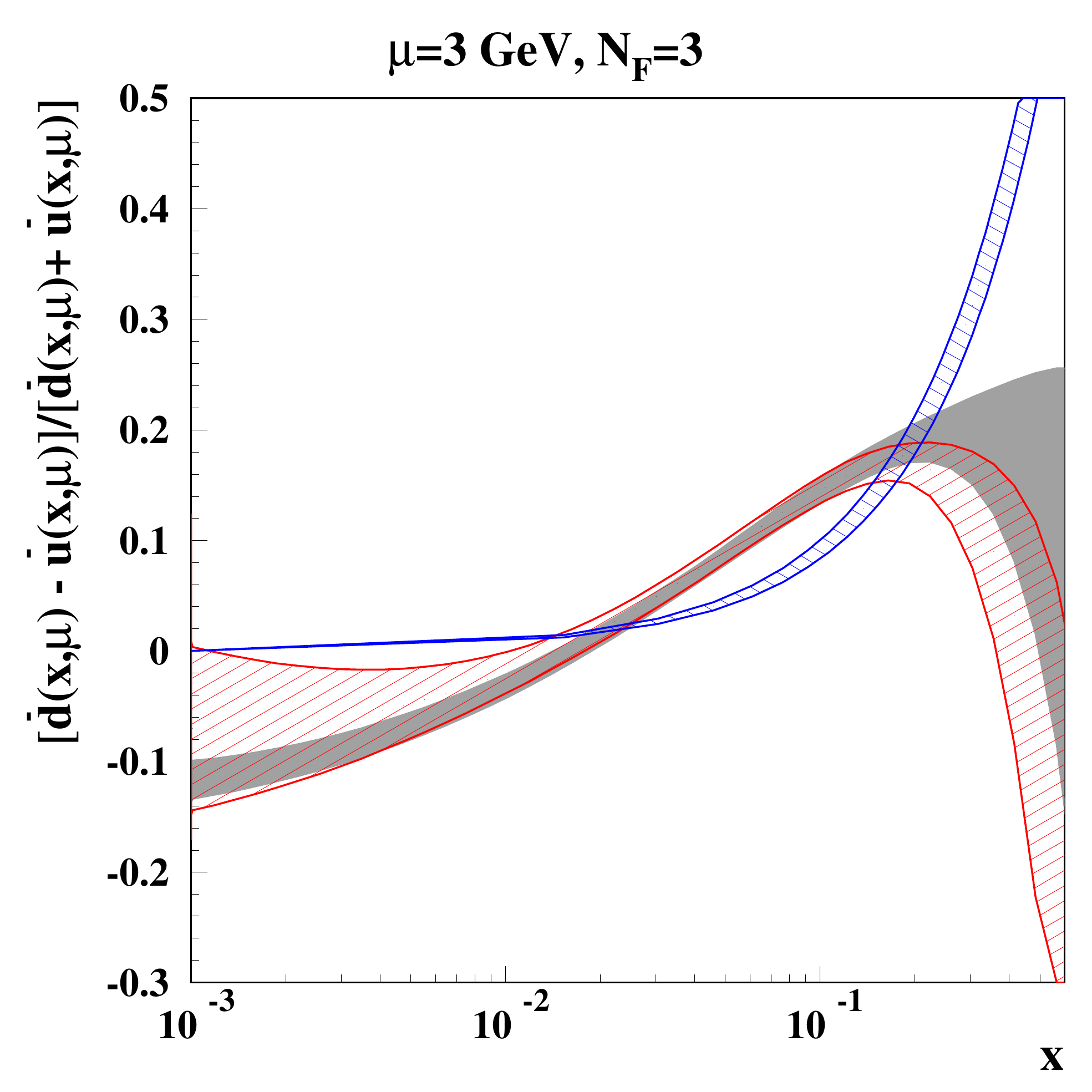}}
  \caption{\small
    \label{fig:udm+ssup1}
    The strangeness suppression factor 
    $r_s(x,\mu^2)$ of Eq.~(\ref{eq:ssup}) (left) and the 
    sea-quark iso-spin asymmetry $I(x,\mu^2)$ of Eq.~(\ref{eq:dbar-ubar})  
    (right) at the scale $\mu=3~{\rm GeV}$ versus $x$ in the $N_F=3$ flavor
    scheme for the nominal ABMP16 PDFs (gray, shaded)  
    and two variants of the ABMP16 fit 
    with DY collider data replaced by deuteron DIS data 
    using the ABMP16 parameterization  in Eq.~(\ref{eq:abmp16-pdfs}) (\textit{\textbf{ABMP16 shape}}, right-tilted hatching)
    and the epWZ16 one in Eq.~(\ref{eq:epWZ16-pdfs}) (\textit{\textbf{epWZ16 shape}}, left-tilted hatching).
  }
\end{figure}

\bigskip

In Fig.~\ref{fig:udm+ssup1} we show the results of these test fits 
for the strangeness suppression factor
\begin{equation}
\label{eq:ssup}
r_s(x,\mu^2) \, = \, \frac{s(x,\mu^2) + \bar s(x,\mu^2)}{\bar d(x,\mu^2) + \bar u(x,\mu^2)}
\, .
\end{equation}
and the sea-quark iso-spin asymmetry 
\begin{equation}
\label{eq:dbar-ubar}
I(x,\mu^2) \, = \, \frac{ \bar d(x,\mu^2) - \bar u(x,\mu^2) }{ \bar d(x,\mu^2) + \bar u(x,\mu^2) }
\, .
\end{equation}
The comparison of the nominal ABMP16 fit in Fig.~\ref{fig:udm+ssup1} 
with the variant \textit{\textbf{ABMP16 shape}}
shows good compatibility of both quantities in the range $x\gtrsim 0.01$. 
This confirms the capability of deuteron data to replace the DY ones in the present study.
However, at smaller values $x\lesssim 0.01$, the uncertainty both in the 
iso-spin asymmetry $I(x)$ and in the strangeness suppression $r_s(x)$ increases
significantly in the variant \textit{\textbf{ABMP16 shape}}. This happens due 
to flexibility of the  PDF parameterization Eq.~(\ref{eq:abmp16-pdfs})
at small $x$, which is determined in the nominal ABMP16 fit by the 
collider DY data relevant for this kinematics and lacks such constraints in 
the \textit{\textbf{ABMP16 shape}} fit. 

In contrast, for the \textit{\textbf{epWZ16 shape}} variant of fit 
with the PDF parameterization of
Eq.~(\ref{eq:epWZ16-pdfs}) the strange sea is enhanced, rising to
about $r_s \sim 0.8$ in the region $x\simeq 0.1$. 
It is worth stressing that no ATLAS data is used in this case. 
It is also interesting that the statistical quality of the
neutrino-induced DIS charm-production data 
description does not deteriorate with the observed strange sea enhancement. 
The total values $\chi^2=167$ and $161$ are obtained in the variants
 \textit{\textbf{epWZ16 shape}} and \textit{\textbf{ABMP16 shape}}, 
respectively, for the combination of 
CCFR/NuTeV, CHORUS, and NOMAD data sets used in the fit. 
This means that the strange sea enhancement is obviously achieved at the expense of a suppressed $d$-quark sea, 
as it was pointed out earlier in Ref.~\cite{Alekhin:2014sya} and is also 
demonstrated in Fig.~\ref{fig:udm+ssup1}.
In addition, the uncertainty band in the iso-spin asymmetry $I(x)$ 
of the \textit{\textbf{epWZ16 shape}} variant of the fit  
is significantly smaller than that for the \textit{\textbf{ABMP16 shape}} and  
for the nominal ABMP16 fit despite the fact that the latter is based on a much wider set of data, even including DY collider data. 

In summary the test variants indicate that the parameterization 
of Eq.~(\ref{eq:epWZ16-pdfs}) underlying the epWZ16 PDFs is over-constrained.
As a matter of fact, Eq.~(\ref{eq:epWZ16-pdfs}) leaves little flexibility for
the iso-spin asymmetry $I(x,\mu^2)$, which can be written as 
\begin{equation}
\label{eq:dbar-ubar-epWZ16}
I(x,\mu^2)\biggr|_{\rm{Eq.~(\ref{eq:epWZ16-pdfs})}} \, = \, A_{\bar{u}}
x^{B_{\bar{u}}} (1-x)^{C_{\bar{u}}} \left( (1-x)^{\delta C} - 1 \right)
\, ,
\end{equation}
where $\delta C = C_{\bar{d}} - C_{\bar{u}}$. 
Depending on the sign of $\delta C$, this function is either positive or
negative definite in the range $0 < x < 1$. 
Thus, it does not allow non-monotonic behaviour of $I(x)$ and, in particular,
a delayed onset of the Regge asymptotics of a vanishing $I(x)$ at
small-$x$ indicated by DY collider data, see also~\cite{Alekhin:2015cza}.

\bigskip

\begin{figure}[t!]
    \centerline{
    \includegraphics[width=11.0cm]{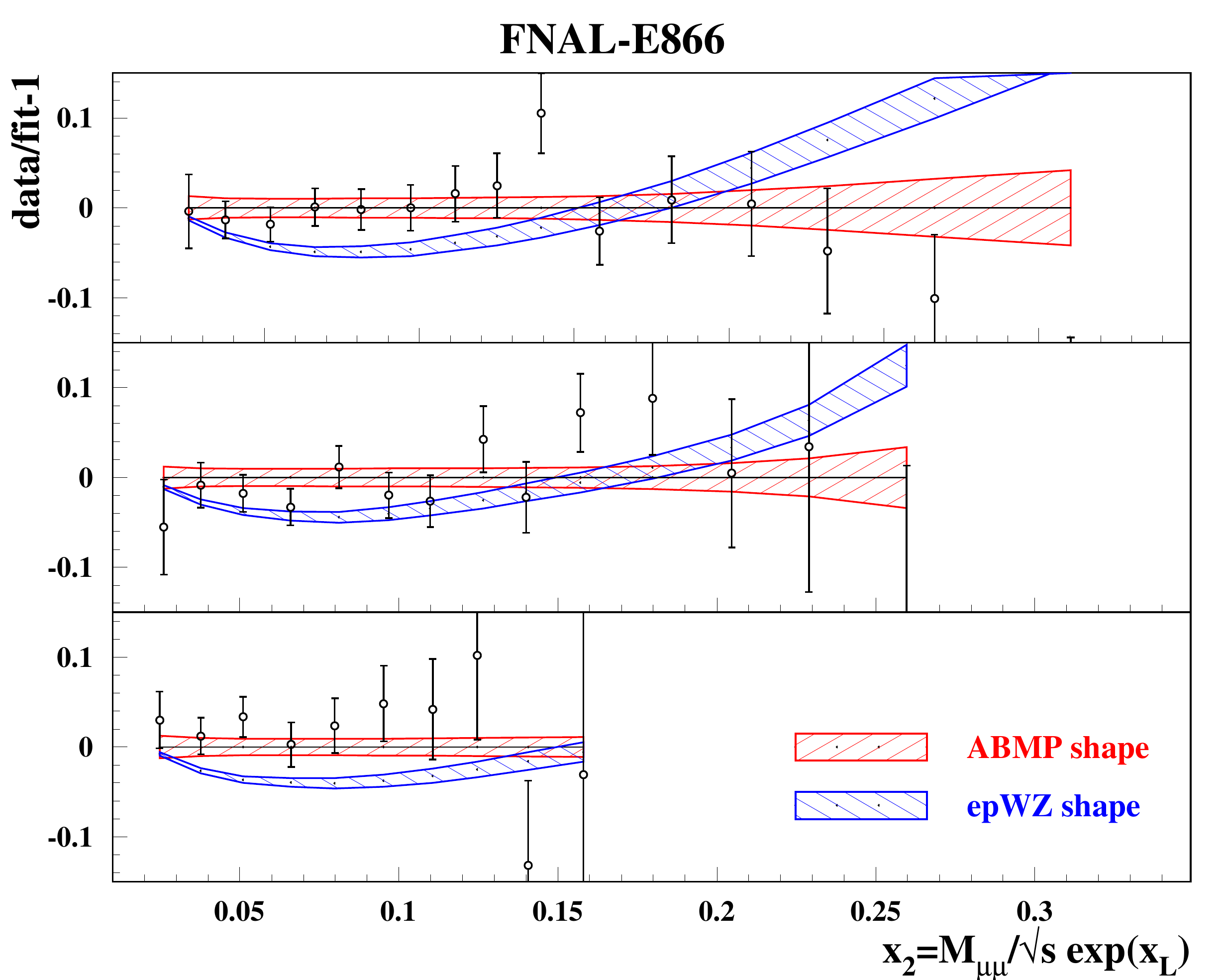}
}
\vspace*{-2mm}
  \caption{\small
    \label{fig:e866-1}
    The pulls of the E866 data on inclusive di-muon production in proton-proton 
    collisions~\cite{Towell:2001nh} for the test fit 
    \textit{\textbf{ABMP16 shape}} versus
    the partonic momentum $x_2=M_{\mu\mu}/\sqrt{s}\,e^{x_L}$, where 
    $M_{\mu\mu}$, $s$ and $x_L$ are invariant mass of the di-muon system, 
    the center-of-mass collision energy and the Bjorken longitudinal   
    momentum of the di-muon system, respectively. 
    The uncertainties in the predictions of the \textit{\textbf{ABMP16 shape}} fit
    (right-tilted hatching) and the difference between its central value and the 
    \textit{\textbf{epWZ16 shape}} fit including 
    uncertainties of the latter (left-tilted hatching) are displayed for 
    comparison.
  }
\vspace*{5mm}
%
\centerline{
  \includegraphics[width=7.85cm]{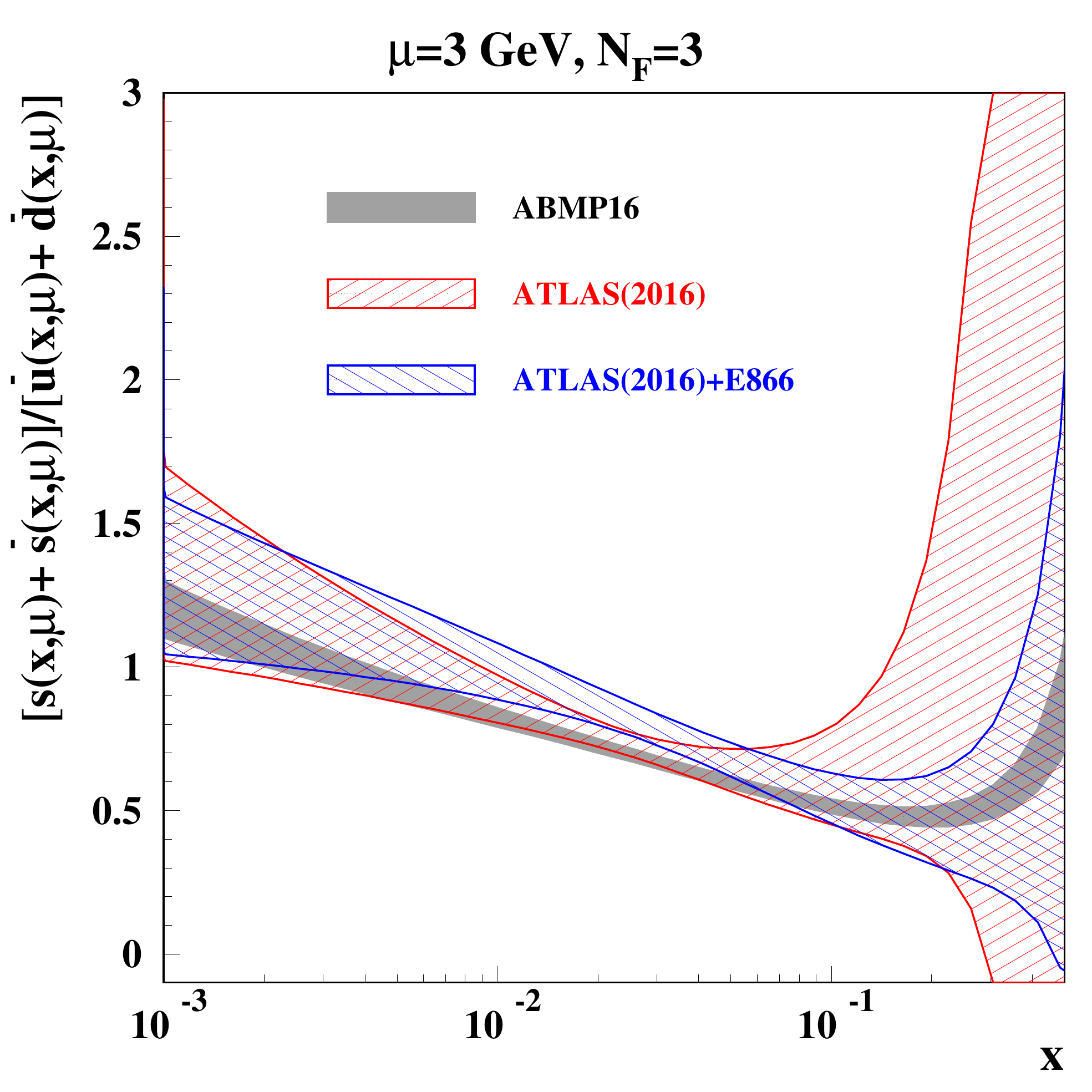}
  \includegraphics[width=7.85cm]{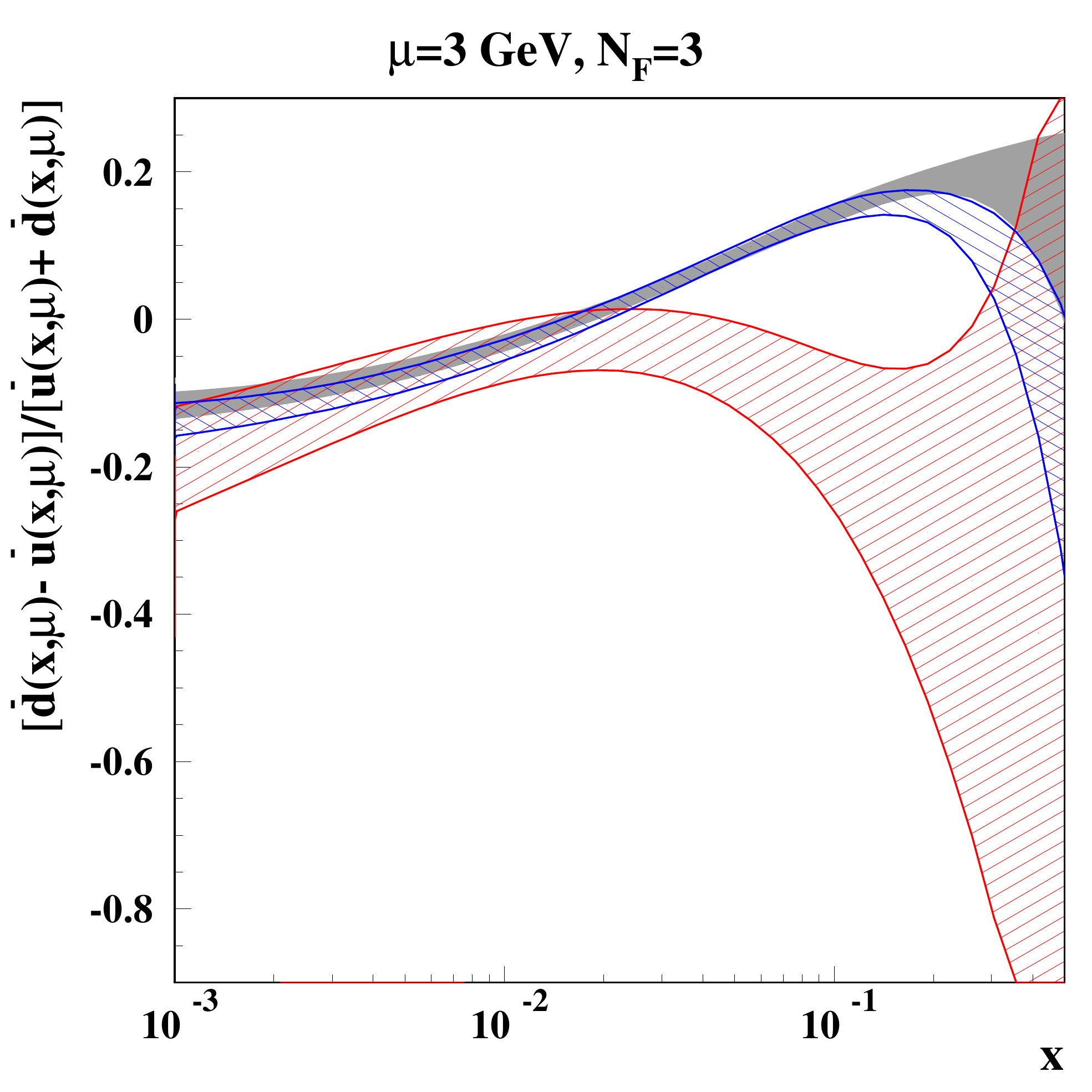}}
\vspace*{-2mm}
  \caption{\small
    \label{fig:udm+ssup2}
  The same as in Fig.~\ref{fig:udm+ssup1}
for the ABMP16 PDFs (gray, shaded) 
    and two variants of the ABMP16 fit with the ATLAS data sets for 
    the $W^\pm \to l^\pm \nu$ and $Z \to l^+l^-$ cross sections from 
    2016~\cite{Aaboud:2016btc}:  
    \textit{\textbf{ATLAS(2016)}} (right-tilted hatching)
    and the same in combination with the fixed-target DY data 
     of the E866 experiment~\cite{Towell:2001nh},
    \textit{\textbf{ATLAS(2016)+E866}} (left-tilted hatching). 
  }
\end{figure}

As a direct consequence of the epWZ16 shape's limited flexibility with respect to
the parameterization of the iso-spin asymmetry 
one observes a poor description of the fixed-target DY data 
collected by the E866 experiment at the beam energy $E_b = 800$~GeV 
and with di-muon invariant masses in the range 4.6$ \le  M_{\mu\mu} \le $12.9~GeV 
in proton-proton and proton-deuteron collisions~\cite{Towell:2001nh}. 
The value of $\chi^2/NDP=96/39$ obtained for this sample in the \textit{\textbf{epWZ16 shape}} fit is about 
twice larger than $\chi^2/NDP=49/39$ for the \textit{\textbf{ABMP16 shape}} fit 
and $\chi^2/NDP=53/39$ for the nominal ABMP16 fit, where $NDP$ denotes the number of data points. 
In Fig.~\ref{fig:e866-1} the \textit{\textbf{epWZ16 shape}} pulls demonstrate a clear off-set at small $x$.
In addition, the uncertainties in the predictions of this fit are greatly 
suppressed, obviously due to the constraint 
$I(x)\to 0$ in the limit $x\to 0$ implied by Eq.~(\ref{eq:dbar-ubar-epWZ16}).

\newpage

\subsection*{3. Combinations of data sets in PDF fits}
In order to check constraints on PDFs due to the combination 
of different data sets we consider two more variants of the ABMP16 fit. 
Specifically, we are interested in the final inclusive $e^\pm p$ scattering cross-section data from H1
and ZEUS~\cite{Abramowicz:2015mha} 
in combination with the ATLAS data~\cite{Aaboud:2016btc} on 
the $W^\pm \to l^\pm \nu$ and $Z \to l^+l^-$ production, 
as these data have been employed in the fit of the epWZ16 PDFs.

To this end, we start from the ABMP16 fit and keep only 
the proton DIS data (see Tab.~2 in Ref.~\cite{Alekhin:2017kpj}), 
while we drop the ones on neutrino-induced charged-current charm-quark production 
(see Tab.~2 in Ref.~\cite{Alekhin:2017kpj}), which have the most essential 
impact on the strange sea determination in the global PDF fits, 
cf. e.g. Ref.~\cite{Alekhin:2014sya}.
Also, all other DY collider data are omitted in order 
to shed light on the potential of the ATLAS sample 
and no deuteron DIS data are used, either.
Such a data selection more or less reproduces the framework of 
the ATLAS analysis~\cite{Aaboud:2016btc} with one exception concerning 
the fixed-target proton data (see Tab.~2 in Ref.~\cite{Alekhin:2017kpj}).
In our case these data are used in order to constrain the large-$x$ PDF behavior.  
Furthermore, we either include or omit the fixed-target DY data of 
the E866 experiment~\cite{Towell:2001nh}, 
which play an essential role in the interpretation 
of the ATLAS data, as we have already discussed earlier~\cite{Alekhin:2014sya}.
Thus, the following two variants are considered:
\begin{itemize}
\item [ ] \textit{\textbf{ATLAS(2016)}} 
  -- 
  a fit based on the proton DIS data used in the ABMP16 analysis and the 2016
   ATLAS data set~\cite{Aaboud:2016btc} for 
  $W^\pm \to l^\pm \nu$ and $Z \to l^+l^-$ production cross sections 
  collected at the collision energy $\sqrt s =7$~TeV with a 
  luminosity of 4.6 fb$^{-1}$
  and cuts on the lepton's transverse momentum $P_T$ of $P_T^l>20~{\rm GeV}$ 
  using the PDF shape of Eq.~(\ref{eq:abmp16-pdfs}). We employ for this study 
  $Z$-boson production data in the central-region, i.e. with the lepton-pair 
  rapidity $\eta_{ll} \lesssim 2.4$, which provide the most accurate  
  data sample of Ref.~\cite{Aaboud:2016btc}.

\item [ ] \textit{\textbf{ATLAS(2016)+E866}} 
  -- the same as the \textit{\textbf{ATLAS(2016)}} fit with
  the E866 data~\cite{Towell:2001nh} added.
\end{itemize}

In Fig.~\ref{fig:udm+ssup2} (left) we observe a somewhat enhanced strange sea from
the variant \textit{\textbf{ATLAS(2016)}} as compared to the nominal ABMP16 fit, 
but within uncertainties the results are well in agreement.
Correspondingly, the \textit{\textbf{ATLAS(2016)}} iso-spin asymmetry at
$x\gtrsim 0.01$ is somewhat smaller than the nominal ABMP16 one, 
as shown in Fig.~\ref{fig:udm+ssup2} (right).
However, again, the discrepancy is statistically not very significant 
due to the quite big uncertainties in the \textit{\textbf{ATLAS(2016)}} variant. 

\begin{figure}[t!]
\centerline{
  \includegraphics[width=11.0cm]{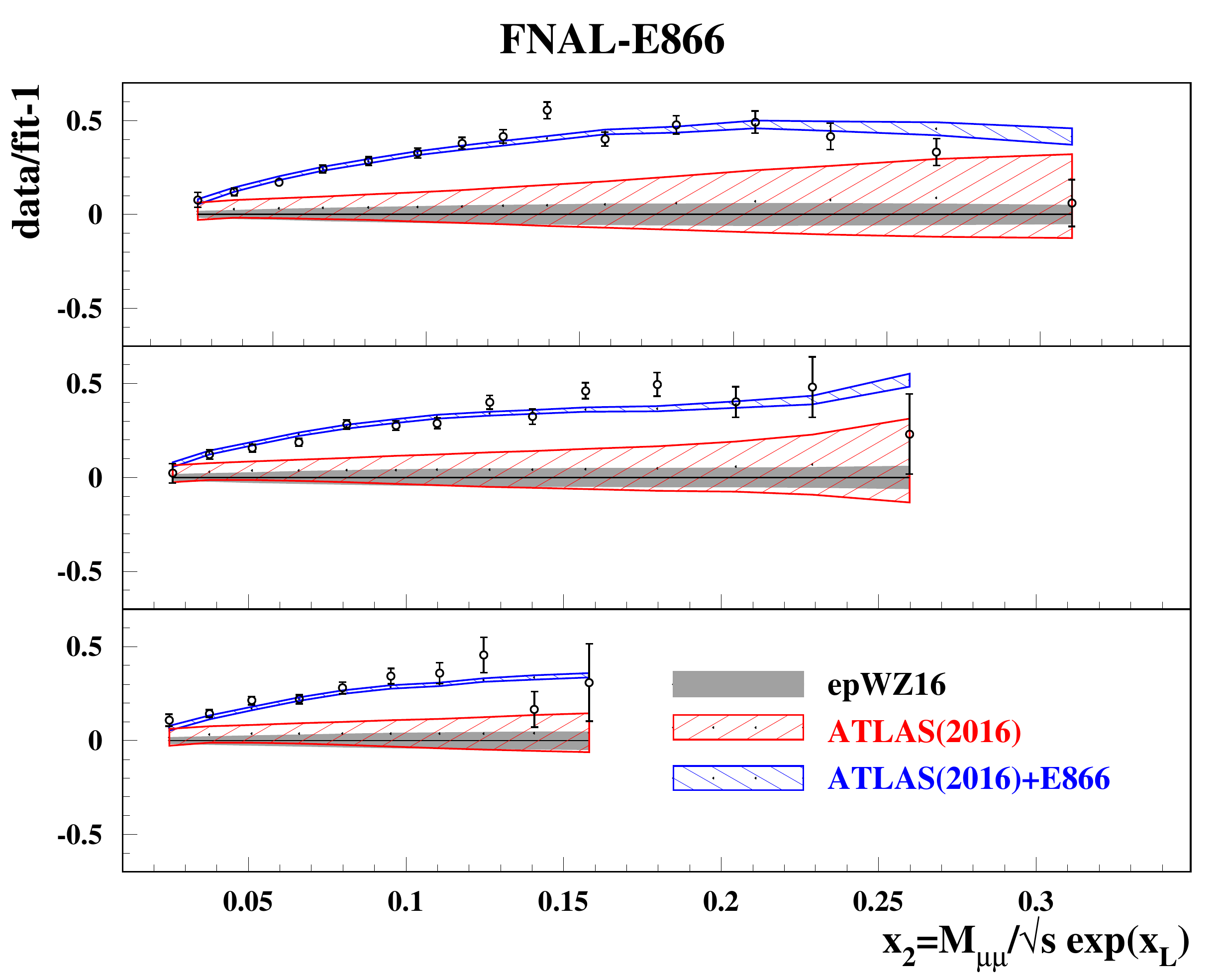}}
  \caption{\small
    \label{fig:e866-2} The same as Fig.~\ref{fig:e866-1} for the 
    prediction of the epWZ16 fit~\cite{Aaboud:2016btc} (gray, shaded) and its 
    difference with predictions of the variants \textit{\textbf{ATLAS(2016)}}
    (right-tilted hatching) and \textit{\textbf{ATLAS(2016)+E866}} (left-tilted hatching).
  }
\vspace*{5mm}
\centerline{
  \includegraphics[width=16cm]{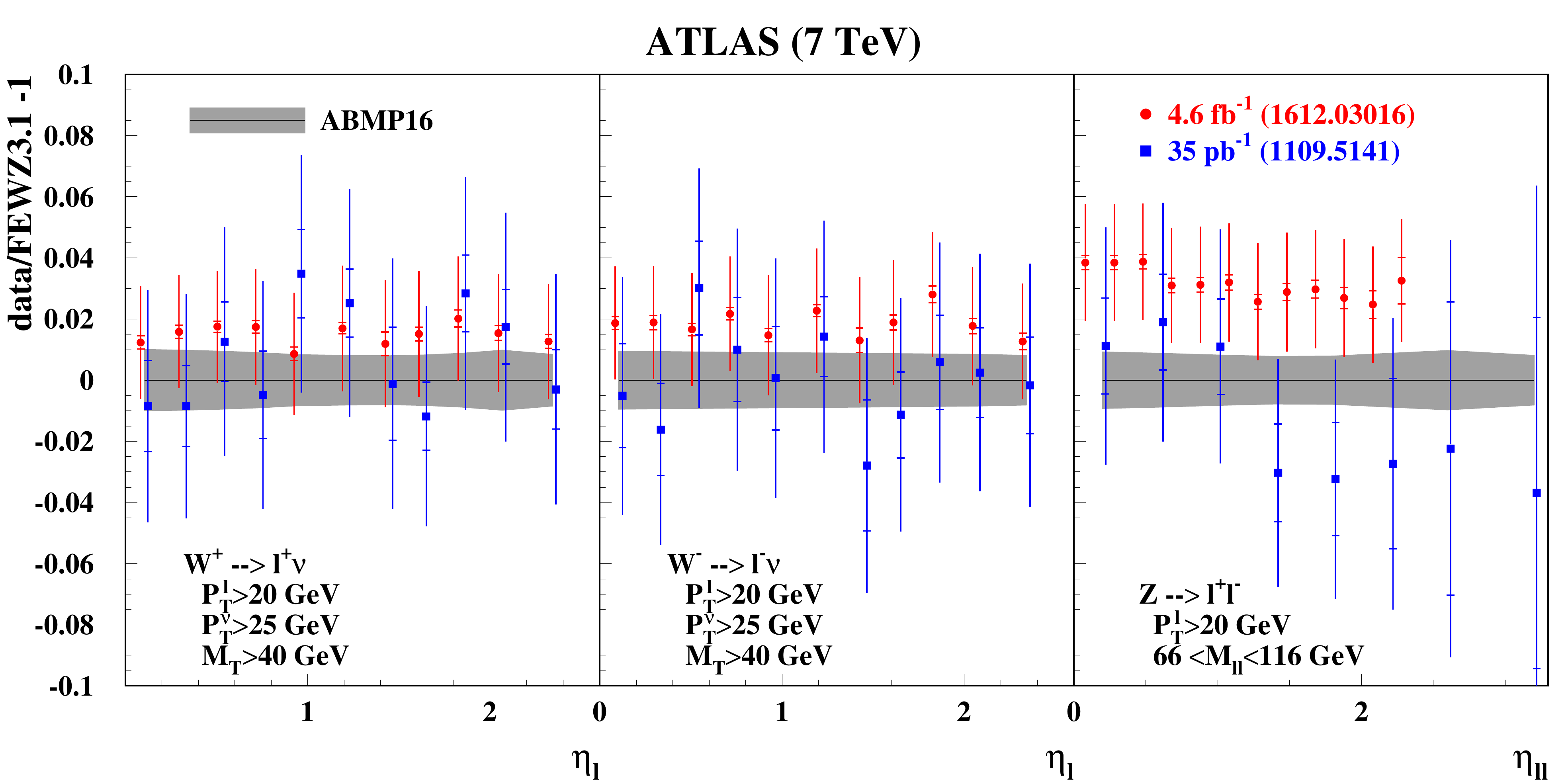}}
  \caption{\small
    \label{fig:WZdata}
    The pulls for the ATLAS data on the 
    $pp \to W^\pm+X \to l^\pm \nu + X$ production (left and center panels)
    and $pp \to Z+X \to l^+l^- + X$ (right panel) at $\sqrt s = 7$~TeV 
    collected at luminosities of 35 pb$^{-1}$ (2011)~\cite{Aad:2011dm} (blue squares)
    and 4.6 fb$^{-1}$ (2016)~\cite{Aaboud:2016btc} (red circles)
    with cuts on the lepton's transverse momentum $P_T^l>20~{\rm GeV}$ 
    as a function of the lepton pseudo-rapidity $\eta_l$ versus NNLO predictions
    obtained using {\tt FEWZ} (version 3.1)~\cite{Li:2012wna,Gavin:2012sy} and the ABMP16 PDFs.
    The uncertainties in predictions (shaded area) are superimposed for 
    comparison. 
  }
\end{figure}

This discrepancy manifests itself also in the comparison with the E866 data, 
which overshoot the \textit{\textbf{ATLAS(2016)}} predictions as shown in Fig.~\ref{fig:e866-2}. 
Since the uncertainties in these predictions are also sizable, 
the E866 data can be well accommodated into the \textit{\textbf{ATLAS(2016)+E866}} variant of fit, 
with  the values of $\chi^2/NDP$=48/39 and 40/34 obtained for the E866 and ATLAS sets, respectively.
When the E866 data is included, the uncertainties both in $I(x)$ and $r_s$
are greatly reduced as displayed already in Fig.~\ref{fig:udm+ssup2}. 
This illustrates in particular the poor potential of the combined HERA and ATLAS data in 
disentangling the light-flavor quark content of the proton.
On the other hand, the value of iso-spin asymmetry $I(x)$ obtained in the 
\textit{\textbf{ATLAS(2016)+E866}} variant is in a good agreement  
with the ABMP16 fit, which also includes the E866 data. 
In view of the clear correlation between strange- and down-quark distributions, the value of 
the strangeness suppression $r_s$ at $x\gtrsim 0.05$ 
in case of the \textit{\textbf{ATLAS(2016)+E866}} variant is smaller than the \textit{\textbf{ATLAS(2016)}} one 
and also in agreement with the ABMP16 result, 
which is driven by the data on neutrino-induced DIS charm-quark production 
to a great extent.  

In summary these comparisons explain the seemingly high precision 
of the strange sea extraction reported in~\cite{Aaboud:2016btc} 
as a direct consequence of using the over-constrained PDF shape Eq.~(\ref{eq:epWZ16-pdfs}). 
Indeed, as we have shown above, cf. Fig.~\ref{fig:udm+ssup1}, 
this shape results in a value of $I(x)$ with underestimated uncertainties. 
In turn, this leads to an enhanced $r_s$ at $x\gtrsim 0.01$, 
which formally appears as being determined with a high accuracy.\footnote{The same trend is observed in the analysis of 
the earlier ATLAS data~\cite{Aad:2011dm}, although in a less pronounced
manner due to less accurate data.} 
In the epWZ16 fit no changes are observed even if the E866 data are included
into the fit~\cite{atlas:private}, 
which can easily be understood since the uncertainties in the epWZ16 predictions
for E866 data are quite small due to the over-constrained PDF shape.
Therefore there is only a little possibility for a variation of the PDFs in direction of the E866 data,
in contrast to the case of the \textit{\textbf{ATLAS(2016)}} fit, cf. Fig.~\ref{fig:e866-2}.

\subsection*{4. ATLAS data vs. ABMP16 PDFs}
The variant \textit{\textbf{ATLAS(2016)+E866}} discussed previously has 
revealed a strange sea suppression factor $r_s$, shown in Fig.~\ref{fig:udm+ssup2}, 
which is somewhat enhanced at $x\sim 0.03$ as compared to the ABMP16 determination.
It is interesting to understand, which ATLAS cross section measurements, i.e., 
for $W^\pm \to l^\pm \nu$ or $Z \to l^+l^-$ is driving this enhancement.
From Fig.~\ref{fig:WZdata} it is obvious, that this happens due to the $Z$-boson production sample, 
which overshoots the ABMP16 prediction.
The ATLAS $Z$-boson data of the 2016 sample is also increased compared to the earlier
data from 2011~\cite{Aad:2011dm} used in the ABMP16 fit, especially in
the region of large lepton-pair rapidity $\eta_{ll} \gtrsim 1$.

Note, that the discrepancy between the prediction of the ABMP16 PDFs and the 
$Z$-boson data in Fig.~\ref{fig:WZdata} is at the level $\sim 1\sigma$. 
Moreover, the 2011 ATLAS data sample~\cite{Aad:2011dm} is well described 
in the ABMP16 fit with $\chi^2/NDP=31/30$ (see Tab.~6 in \cite{Alekhin:2017kpj}).
Therefore, the new high statistics ATLAS data can evidently be accommodated 
into the ABMP16 fit with a reasonable value of $\chi^2$ and at the price 
of a moderate variation of the PDFs.

\begin{figure}[t!]
\centerline{
  \includegraphics[width=10cm]{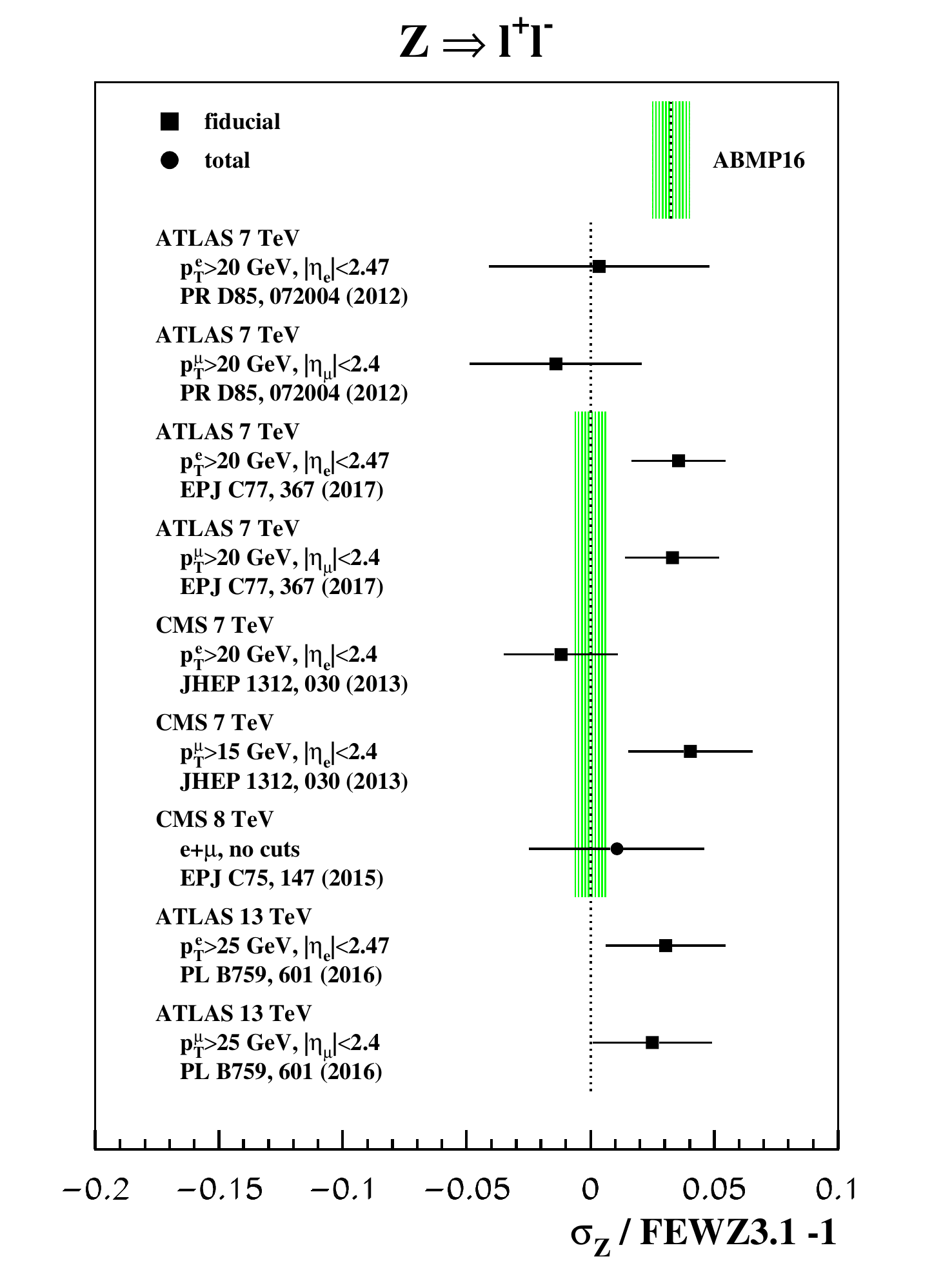}}
  \caption{\small
    \label{fig:zsec}
    The same as Fig.~\ref{fig:WZdata} for the integrated cross sections of $Z$-boson production in 
    proton-proton collisions in the central-region measured by the ATLAS and CMS experiments in the 
    $e$- and $\mu$-decay channels at different center-of-mass 
    energies~\cite{Aaboud:2016btc,Aad:2011dm,Chatrchyan:2013tia,CMS:2014jea,Aad:2016naf}.
    The predictions for those data sets, which 
    are not included into the ABMP16 analysis, are displayed with uncertainties (shaded area).
  }
\end{figure}

In order to clarify the robustness of such a potential variation of the PDFs, 
we compare the new ATLAS results~\cite{Aaboud:2016btc} with other ATLAS and CMS data on $Z$-boson production. 
For this comparison we select the integrated cross section 
for the central production with rapidities $\eta_{ll} \lesssim 2.4$ 
measured at the center-of-mass energies $\sqrt{s}=7$, 8 and 13~TeV~\cite{Aad:2011dm,Chatrchyan:2013tia,CMS:2014jea,Aad:2016naf}.
The kinematics of these measurements are similar to the one of ATLAS~\cite{Aaboud:2016btc}, 
so that they allow to understand how the latter data compare to the general trend. 
The ABMP16 predictions at NNLO in QCD  are in agreement with the earlier ATLAS data at $\sqrt{s}=7$~TeV~\cite{Aad:2011dm}, 
which were used in the fit, while the recent ones~\cite{Aaboud:2016btc}
somewhat overshoot the predictions and the $\sqrt{s}=13$~TeV sample demonstrate similar trend, cf. Fig.~\ref{fig:zsec}.
At the same time the latest ATLAS data are on average about $\sim 2$\% higher than
the CMS ones at $\sqrt{s}=7$ and 8~TeV. 
This difference, although being with the experimental uncertainties, indicates 
that the CMS data prefer strange sea distribution comparable with the one in the ABMP16 PDFs. 

\begin{figure}[t!]
\centerline{
  \includegraphics[width=10cm]{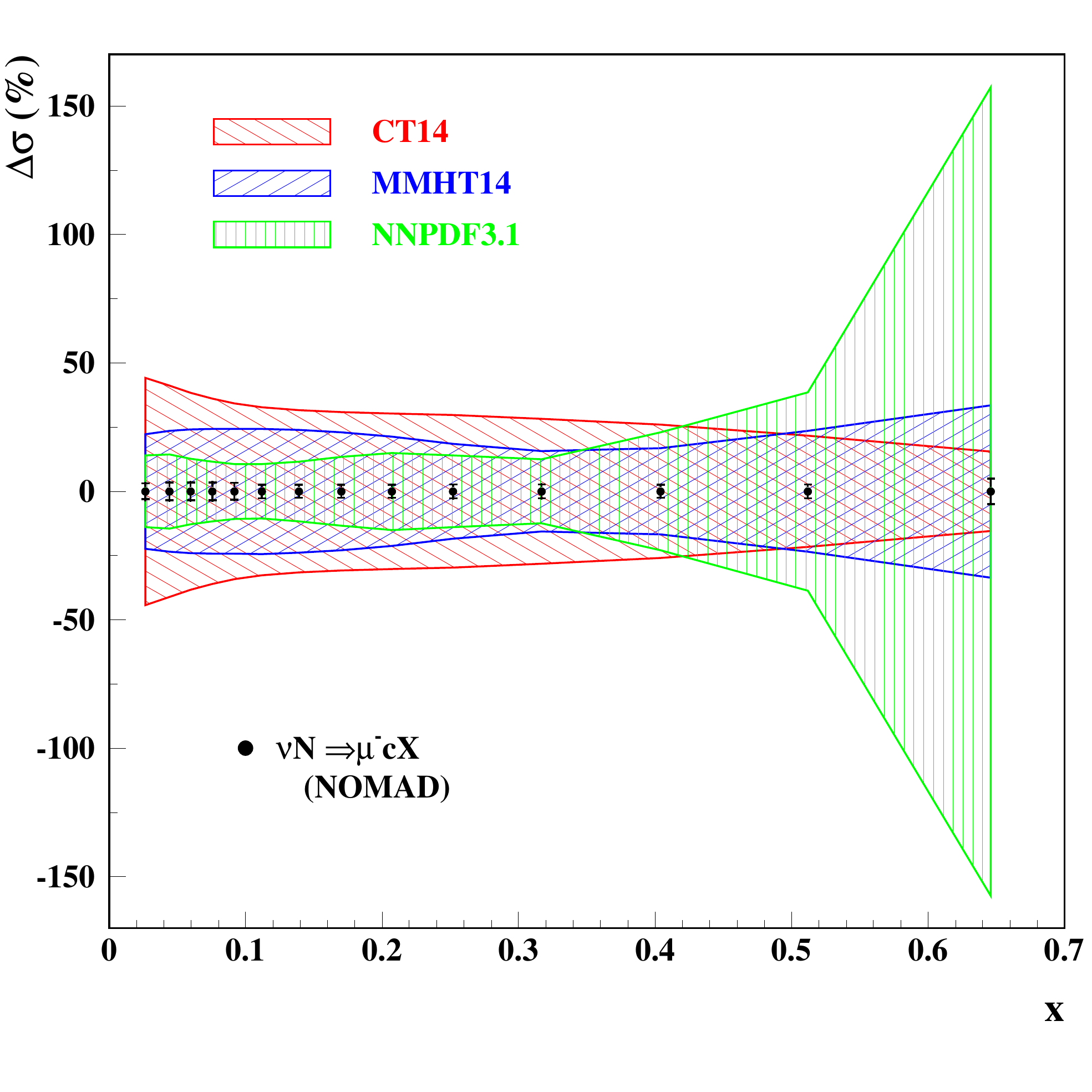}}
  \caption{\small
    \label{fig:nomad}
    The uncertainties in the NOMAD data on the cross sections of
    charged-current neutrino-induced 
    DIS charm-quark production~\cite{Samoylov:2013xoa} versus Bjorken $x$
    in comparison to
    predictions of various PDFs:  
    CT14~\cite{Dulat:2015mca} (left-tilted hatch), 
    MMHT14~\cite{Harland-Lang:2014zoa} (right-tilted hatch), 
    NNPDF3.1~\cite{Ball:2017nwa} (vertical hatch).  
  }
\end{figure}

\subsection*{5. Outlook}
Having clarified the constraints on the strange sea PDFs from the 
recent ATLAS data on $W^\pm$- and $Z$-boson production and 
having scrutinized analysis details underlying the determination of the epWZ16 PDFs 
we would like to close by reviewing prospects for further improvements in the
precision of the strangeness distribution in the proton. 

First of all, there is the associated production of $W^\pm$-bosons and a
charm-quark which has been measured both by ATLAS and CMS~\cite{Aad:2014xca,Chatrchyan:2013uja}.
These data had already been analyzed regarding their impact on the strange
PDFs in \cite{Alekhin:2014sya}. In order to improve upon the current status, 
measurements of the lepton rapidity distributions $d\sigma(W^\pm+c)/d\eta_l$ 
for the individual $W^+ \bar{c}$ and $W^- {c}$ channels with an ${\cal O}(3\%)$ accuracy are needed.
The other limiting factor is the lack of the complete NNLO QCD corrections for
those differential $d\sigma(W^\pm+c)/d\eta_l$ cross sections.

\bigskip

In the meantime, other global fits can be improved with the help of the NOMAD data 
on charged-current neutrino-induced DIS charm-quark production~\cite{Samoylov:2013xoa}.
The impact of the NOMAD data on the determination of the strange sea at medium $x$ 
is illustrated in Fig.~\ref{fig:nomad} by comparing the experimental uncertainties in the data with the 
ones of the predictions based on various PDF fits.
In Fig.~\ref{fig:nomad} we select the sets of PDFs by CT14~\cite{Dulat:2015mca}, 
MMHT14~\cite{Harland-Lang:2014zoa} and NNPDF3.1~\cite{Ball:2017nwa}, 
all of which have been fitted to a bulk of global data, however, not including
those by the NOMAD experiment.

As a result of this lack of constraints from data, the uncertainties 
in the predictions by CT14, MMHT14 or NNPDF3.1 reach ${\cal O}(20\%)$, cf. Fig.~\ref{fig:nomad}. 
For the case of NNPDF3.1 at large-$x$, $x \gtrsim 0.5$, the uncertainties are even larger than ${\cal O}(100\%)$, 
which implies that some of the PDF replica predict negative cross sections.\footnote{
  NNPDF3.1~\cite{Ball:2017nwa} has also performed a fit variant based on the same data as the epWZ16 fit
  of ATLAS~\cite{Aaboud:2016btc} and finds that strangeness is significantly enhanced.
  This is possible due to the phenomenological model for the
  charm-quark PDF applied in~\cite{Ball:2017nwa}, which considers the entire charm PDF as a
  function to be fitted, 
  rather than considering charm to be produced dynamically as predicted 
  by QCD~\cite{Witten:1975bh,Laenen:1992zk,Ablinger:2016swq}.}
Such unphysical features also cast severe doubt on the averaging procedure performed
on the basis of these three PDF sets as advocated by the PDF4LHC 
recommendation~\cite{Butterworth:2015oua}.
Note, that the NOMAD data have much smaller uncertainties, $\sim~5$\%, 
and therefore can evidently help to consolidate the spread in predictions 
providing a more accurate framework for the precision studies, 
like the $W$-boson mass measurement~\cite{Aaboud:2017svj}. 
Moreover, they are presented in the form of the dimuon and total charged-current
 cross section ratio, which is much less sensitive to the impact of 
nuclear corrections, as compared to the CCFR and NuTeV data.

\bigskip

Lattice QCD simulations may, in the future, also help to determine strangeness in the proton.
The fraction of the nucleon's momentum carried by the strange quark has
recently been extracted from nucleon matrix elements of suitable operators 
at the physical pion mass~\cite{Alexandrou:2017oeh}, 
but the current value still carries a rather large uncertainty of ${\cal O}(50\%)$.

\bigskip

Finally, due to the documented correlation between the strange sea quarks and
the non-strange ones, it will be important to measure the ${\bar u}$- and
${\bar d}$-PDFs in the proton, specifically the ratio ${\bar d}/{\bar u}$ 
at large $x \gtrsim 0.1$.
Here, Fermilab's E906 SeaQuest experiment can help to
improve the situation by measuring muons in the DY process from proton-proton and
proton-deuterium scattering. These data can be used to extract the 
ratio ${\bar d}/{\bar u}$ for the light-quark sea at Bjorken-$x$ values up to
$x \simeq .45$, i.e. to higher $x$ than in previous E866 experiment~\cite{seaquest:2017}.

{\bf{Acknowledgements:}}\qquad 
We would like to thank A.~Glazov and U.~Klein for discussions regarding the 
ATLAS data and details of the analysis in Ref.~\cite{Aaboud:2016btc}. 
This work has been supported by Bundesministerium f\"ur Bildung und Forschung (contract 05H15GUCC1)
and by the European Commission through PITN-GA-2012-316704 ({\it HIGGSTOOLS}). 

\bigskip

\providecommand{\href}[2]{#2}\begingroup\raggedright\endgroup

\end{document}